\documentclass[aps,prl,twocolumn,floatfix,altaffilletter,superscriptaddress,preprintnumbers,
tightenlines,showpacs,showkeys,nofootinbib,notoccite]{revtex4-1}
\usepackage[utf8]{inputenc}
\usepackage[colorlinks=true,citecolor=blue,linkcolor=blue]{hyperref}
\usepackage[normalem]{ulem}
\usepackage{amsmath,amssymb, mathrsfs,esvect}
\usepackage{epsfig}
\usepackage{graphicx}               
\usepackage{url}
\usepackage{color}
\usepackage{slashed}
\usepackage{multirow}
\usepackage{placeins}
\usepackage[dvipsnames]{xcolor}
\usepackage{epstopdf}
\usepackage{soul}
\usepackage{tikz}
\usepackage[capitalise, english]{cleveref}
\usepackage{siunitx}
\usepackage{xspace}
\usepackage{booktabs}
\usetikzlibrary{trees}
\usetikzlibrary{decorations.pathmorphing}
\usetikzlibrary{decorations.markings}

\newcommand\myshade{80}
\colorlet{mylinkcolor}{Blue}
\colorlet{mycitecolor}{Red}
\colorlet{myurlcolor}{violet}

\newcommand{\sigv}{\langle \sigma v \rangle_{\rm ann}}

\newcommand{\beq}{\begin{eqnarray}}
\newcommand{\eeq}{\end{eqnarray}}

\hypersetup{
	linkcolor  = mylinkcolor!\myshade!black,
	citecolor  = mycitecolor!\myshade!black,
	urlcolor   = myurlcolor!\myshade!black,
	colorlinks = true
}

\allowdisplaybreaks

\usepackage{tikz,xcolor,hyperref}

\definecolor{lime}{HTML}{A6CE39}
\DeclareRobustCommand{\orcidicon}{\hspace{-1mm}
	\begin{tikzpicture}
		\draw[lime, fill=lime] (0,0) 
		circle [radius=0.15] 
		node[white] {{\fontfamily{qag}\selectfont \tiny \,ID}};
		\draw[white, fill=white] (-0.0525,0.095) 
		circle [radius=0.007];
	\end{tikzpicture}
	\hspace{-2mm}
}

\foreach \x in {A, ..., Z}{\expandafter\xdef\csname orcid\x\endcsname{\noexpand\href{https://orcid.org/\csname orcidauthor\x\endcsname}
		{\noexpand\orcidicon}}
}
\keywords{}

\begin{document}

\title{ Dark Matter Annihilation inside Large Volume Neutrino Detectors}

\author{David McKeen}
\email{mckeen@triumf.ca}
\affiliation{TRIUMF, 4004 Wesbrook Mall, Vancouver, BC V6T 2A3, Canada}

\author{David E. Morrissey}
\email{dmorri@triumf.ca}
\affiliation{TRIUMF, 4004 Wesbrook Mall, Vancouver, BC V6T 2A3, Canada}

\author{Maxim Pospelov}
\email{pospelov@umn.edu}
\affiliation{School of Physics and Astronomy, University of Minnesota, Minneapolis, MN 55455, USA}
\affiliation{William I. Fine Theoretical Physics Institute, School of Physics and Astronomy, University of Minnesota, Minneapolis, MN 55455, USA}

\author{Harikrishnan Ramani}
\email{hramani@stanford.edu}
\affiliation{Stanford Institute for Theoretical Physics, Stanford University, Stanford, California 94305, USA}

\author{Anupam Ray\orcidE{}}
\email{anupam.ray@berkeley.edu}
\affiliation{Department of Physics, University of California Berkeley, Berkeley, California 94720, USA}
\affiliation{School of Physics and Astronomy, University of Minnesota, Minneapolis, MN 55455, USA}

\date{\today}


\begin{abstract}
New particles in theories beyond the Standard Model can manifest as stable relics that interact strongly with visible matter and make up a small fraction of the total dark matter abundance. Such particles represent an interesting physics target since they can evade existing bounds from direct detection due to their rapid thermalization in high-density environments. In this work we point out that their annihilation to visible matter inside large-volume neutrino telescopes can provide a new way to constrain or discover such particles. The signal is the most pronounced for relic masses in the GeV range, and can be efficiently constrained by existing Super-Kamiokande searches for di-nucleon annihilation. We also provide an explicit realization of this scenario in the form of secluded dark matter coupled to a dark photon, and we show that the present method implies novel and stringent bounds on the model that are complementary to direct constraints from beam dumps, colliders, and direct detection experiments.
\end{abstract}

\maketitle
\preprint{N3AS-23-007}


\noindent\emph{\textbf{Introduction:}} Cosmological observations provide nearly unambiguous evidence for a non-baryonic form of matter, commonly known as dark matter (DM), as a dominant component of the Universe~\cite{Aghanim:2018eyx}. Despite extensive searches, the microscopic identity of DM is yet to be revealed. In the absence of a convincing signal thus far, terrestrial and astrophysical searches have placed stringent constraints on the non-gravitational interactions of DM  over a wide mass range~\cite{Cooley:2022ufh,Baryakhtar:2022hbu,Boddy:2022knd}.

While DM might consist of just a single new particle, it could also be composed of several. Indeed, many theories of new physics beyond the Standard Model (SM) predict one or more stable particles, each of which could contribute to the total density of DM. An intriguing example is a new species $\chi$ that interacts strongly with ordinary matter (in the sense of large interaction cross sections and not necessarily the \emph{strong force}) but that makes up only a tiny fraction $f_\chi = \rho_\chi/\rho_\text{DM} \ll 1$ of the total DM mass density. Such relics might seem easy to detect in existing laboratory searches for DM through their scattering with nuclear targets, but they turn out to be much more elusive, see {\em e.g.}, Refs.~\cite{Farrar:2002ic,Collar:2018ydf}. This is simply because a strongly interacting DM component would be slowed significantly by scattering with matter in the atmosphere or the Earth before reaching the target, leading to energy depositions in the detector that are too small to be observed with standard methods~\cite{Zaharijas:2004jv}.

Owing to their interactions with ordinary matter, a strongly-interacting dark matter component (DMC) would be trapped readily in the Earth and thermalize with the surrounding matter.
Furthermore, for lighter DM, strong matter interactions allow Earth-bound DM particles to distribute more uniformly over the entire volume of the Earth rather than concentrating near the center. Together, this can make the DM density near the surface of the Earth tantalizingly large, up to  $\sim f_\chi\times 10^{15}\,\rm{cm}^{-3}$ for DM mass of 1\,GeV~\cite{Neufeld:2018slx,Pospelov:2020ktu,Leane:2022hkk,Berlin:2023zpn}. Despite their large surface abundance, such thermalized DMCs are almost impossible to detect in traditional direct detection experiments as they carry a minuscule amount of kinetic energy $\sim kT = 0.03\,\rm{eV}$. A few recent studies have proposed searches for such a trapped DMC fraction via up-scattering through nuclear isomers~\cite{Pospelov:2019vuf, Lehnert:2019tuw}, electric field acceleration~\cite{Pospelov:2020ktu} and collisions~\cite{McKeen:2022poo}, via bound state formation~\cite{Berlin:2021zbv}, and by utilizing low threshold quantum sensors~\cite{Budker:2021quh, Das:2022srn,Billard:2022cqd}. 

In this work, we propose a novel detection scheme for a GeV-scale DMC $\chi$  with matter fraction $f_\chi \ll 1$ and a large effective scattering cross section with nucleons $\sigma_{\chi n}\gtrsim 10^{-34}\,\text{cm}^2$. The scheme is based on the direct annihilation of the Earth-bound population of DMCs within the active volumes of large neutrino telescopes. As annihilation releases up to $2m_\chi$ of visible energy, it naturally provides a dramatic signal for detection of the relic. Currently, the Super-Kamiokande~(SK) experiment, owing to its enormous fiducial volume and relatively low detection energy threshold, provides the most stringent probe of Earth-bound DMCs via annihilation. We demonstrate that Earth-bound DMC particles in the mass range of $\sim (1-5)$\,GeV can be efficiently constrained via their local annihilation at SK. The lower end of the mass range is determined by the finite temperature of the Earth, whereas, the upper end is set primarily by the gravitational suppression of the surface density of the $\chi$ particles. A similar scheme for direct annihilation inside large volume detectors has previously been discussed for the case of millicharged DM particles \cite{Pospelov:2020ktu}. To illustrate the power of the method within a specific model, we apply it to secluded dark matter that connects to the SM through a dark photon~\cite{Pospelov:2007mp}, and derive new constraints on the parameter space.

\begin{figure*}
	\centering
	\includegraphics[width=0.45\textwidth]{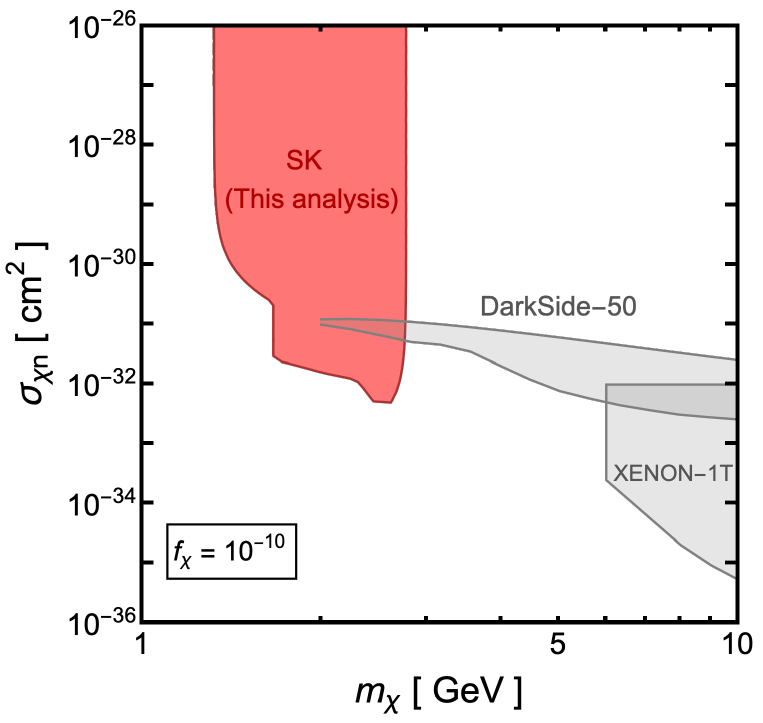}
	\hspace*{1.0 cm}
	\includegraphics[width=0.45\textwidth]{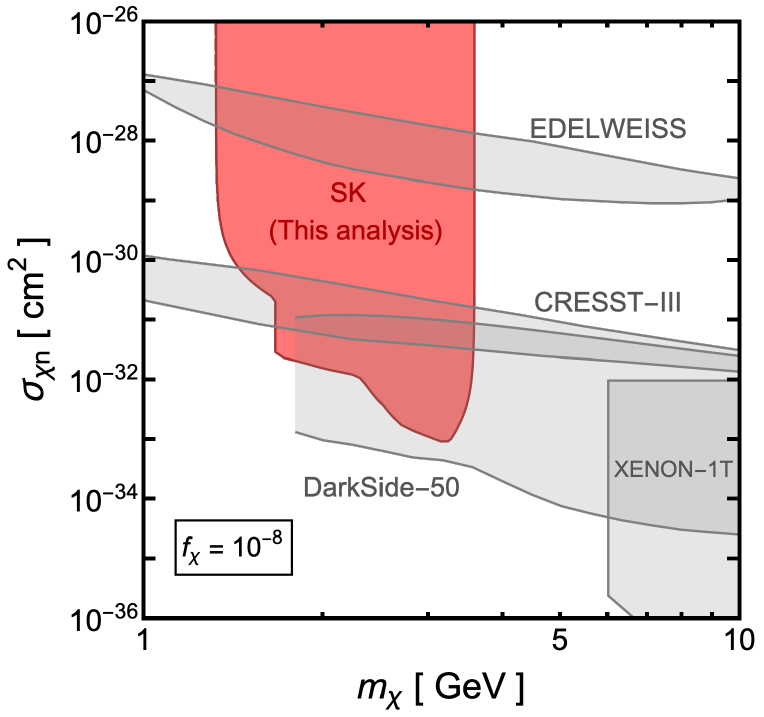} \\
	\includegraphics[width=0.45\textwidth]{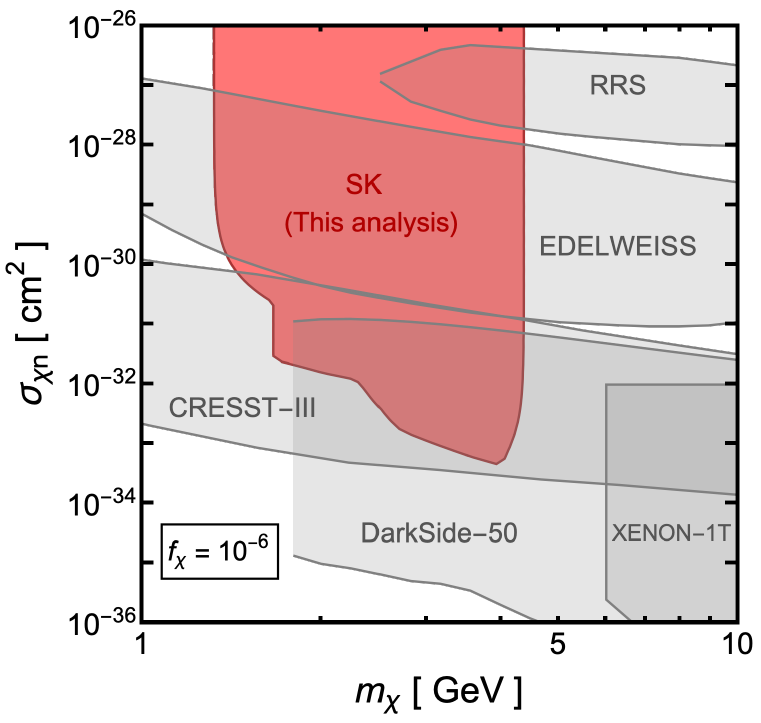}
	\hspace*{1.0 cm}
	\includegraphics[width=0.45\textwidth]{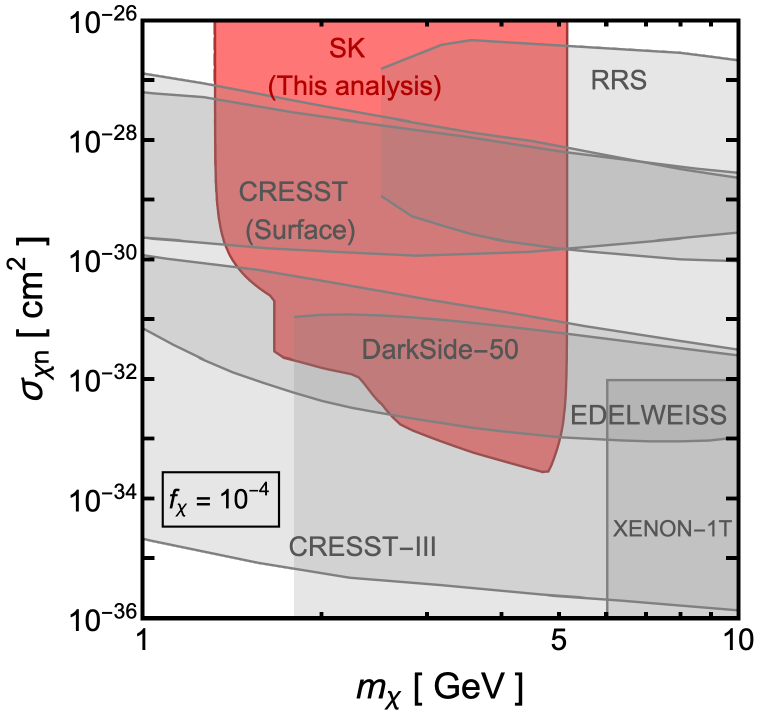}
	\caption{Expected constraints on the DM-nucleon scattering cross-section $\sigma_{\chi n}$ from non-observation of DMC annihilation  inside the fiducial volume of Super-Kamiokande~(red shaded). Each panel shows a specific mass fraction $f_\chi$: $f_\chi = 10^{-10}$~(top left), $f_\chi = 10^{-8}$~(top right), $f_\chi = 10^{-6}$~(bottom left), $f_\chi = 10^{-4}$~(bottom right). For comparison we also show the estimated constraints from direct detection experiments including CRESST~III~\cite{CRESST:2019jnq}, CRESST surface~\cite{CRESST:2017ues}, XENON~\cite{XENON:2018voc}, EDELWEISS surface~\cite{EDELWEISS:2019vjv}, RRS~\cite{Rich:1987st}, and Darkside-50~\cite{DarkSide:2018bpj}~(gray shaded).}
	\label{fig1}
\end{figure*}

\noindent \emph{\textbf{Accumulation and Distribution of DMC:}}
Consider a DMC $\chi$ with mass $m_\chi$, DM fraction $f_\chi$, effective nucleon cross section $\sigma_{\chi n}$, and self-annihilation cross section $\sigv$. If the relic density of $\chi$ arises from thermal freeze-out, the fraction $f_\chi$ can be determined from the annihilation rate in the early Universe with an approximate relation $f_\chi \propto 1/\sigv(T\simeq m_\chi/25)$. Extrapolating this high-temperature cross section to the present-day terrestrial environment depends in a crucial way on the underlying microphysics. In what follows we will concentrate for the most part on $s$-wave annihilation, which implies a nearly constant $\langle \sigma v \rangle_{\rm ann}$. 

The total number of $\chi$ particles $N_\chi$ inside the Earth evolves as
\begin{align}
\label{Nchi}
\frac{dN_\chi}{dt} = \Gamma_{\rm cap} - N_\chi\tau^{-1}_{\rm evap} - N_\chi^2\tau^{-1}_{\rm ann} \ ,
\end{align}
The right hand side of this equation contains the capture, evaporation and annihilation rates; we will discuss each of them in detail below. If dynamical equilibrium is reached, $dN_\chi/dt=0$.

Starting with the capture rate $\Gamma_{\rm cap}$, we can write it as 
\begin{align}
\Gamma_{\rm cap}  =f_{\rm cap} \times \Gamma_{\rm{geom}} =f_{\rm cap} \times \sqrt{\frac{8}{3 \pi}} \frac{f_{\chi} \rho_{\rm DM}v_{\rm gal}}{m_{\chi}} 
\times \pi R^2_{\oplus} \,,
\end{align}
where $\rho_{\rm DM} = 0.4\,\rm{GeV}\,\rm{cm}^{-3}$ denotes the local Galactic DM density, $v_{\rm gal} = 220$ km/s is the typical velocity of the DM particles in the Galactic halo, and $R_{\oplus}$ is the radius of the Earth. We have also defined here the \textit{geometric capture rate} $\left(\Gamma_{\rm{geom}}\right)$, which occurs when all the $\chi$ particles that impact the Earth get trapped. The quantity $f_{\rm cap}$ denotes the capture fraction that accounts for deviations from the geometric rate; for strongly-interacting DMCs, for which the Earth is optically thick, $f_{\rm cap}$ depends on the relic mass. It approaches unity for $m_\chi \gg m_A$, where $m_A$ is a typical nuclear mass in the Earth, while lighter DMCs have a reduced $f_{\rm cap}$ due to reflection. We use the recent numerical simulations of Ref.~\cite{Bramante:2022pmn} to estimate the value of $f_{\rm cap}$, which are found to agree reasonably well with previous analytical estimates~\cite{Neufeld:2018slx}; for $m_\chi = 1$~GeV we find $f_{\rm cap}\simeq 0.1$.

In order to determine $\tau^{-1}_{\rm evap} $ and $ \tau^{-1}_{\rm ann}$, we need to address the spatial distribution of the Earth-bound DM particles inside the Earth. To this end, we introduce the number density of captured $\chi$ particles, $n_\chi(r)$, along with the dimensionless radial profile function, $G_\chi(r)$,
\begin{align}
\int^{R_{\oplus}}_{r=0}\!dr\, 4 \pi r^2 n_{\chi} (r) = N_{\chi}, ~~G_\chi(r) \equiv \frac{V_\oplus n_\chi}{N_\chi} \ . 
\end{align}
For the uniform, radius independent, distribution of $\chi$, the profile function is trivial, $G_\chi(r) =1$. 
To determine $n_\chi(r)$, one turns to the Boltzmann equation that combines the effects of gravity,  concentration diffusion, and thermal diffusion~\cite{Gould:1989hm,Leane:2022hkk}. Moreover, noting that the diffusional timescales for $\chi$ particles are short compared with all other scales in the problem, one can use the hydrostatic equilibrium equation 
\begin{align}
	\frac{\nabla n_{\chi}(r) }{n_{\chi}(r)}+ \left(\kappa +1 \right) 	\frac{\nabla T(r) }{T(r)}+\frac{m_{\chi} g(r)}{k_BT(r)} = 0 \ 
	\label{2}
\end{align}
where $T(r)$ denotes the temperature profile of the Earth and $g(r)$ is its density profile, which we obtain from Refs.~\cite{Dziewonski:1981xy,https://doi.org/10.1002/2017JB014723}. The coefficient responsible for thermal diffusion, $\kappa \sim -1/\left[2(1+m_{\chi}/m_{A})^{3/2}\right]$, is independent of $\sigma_{\chi n}$ as long it remains approximately constant within the range of thermal energies. Rescaling to write this expression in terms of $G_{\chi}(r)$, it is, importantly, independent of the total number of trapped particles $N_\chi$.
Upon solving Eq.~\eqref{2}, we find that for $m_\chi\lesssim 5~\rm GeV$ the density profile is relatively constant and increases only mildly toward the Earth's center. For larger $m_\chi$, the $\chi$ particles tend to settle toward the core and have much smaller density near the surface.

Evaporation is particularly important for light DMCs because thermal processes within the Earth can give sufficient amount of energy to the particles for escape.. In the optically thick regime, evaporation of strongly interacting DMCs is impeded by their scattering with material in the Earth and the atmosphere on the way out~\cite{Neufeld:2018slx}. We adopt the Jeans' expression for the evaporation rate in this regime~\cite{Neufeld:2018slx}, 
\begin{align}
\tau_{\rm evap}^{-1} = G_\chi(R_{\rm LSS})\times \frac{3R_{\rm LSS}^2}{R_\oplus^3}\times \frac{v_{\rm LSS}^2+v_{\rm esc}^2}{2\pi^{1/2}v_{\rm LSS}} \exp\left(- \frac{v_{\rm esc}^2}{v_{\rm LSS}^2}\right),
\end{align}
where $R_{\rm LSS}$ and $v_{\rm LSS}$ are the radius and DM thermal velocity at the last scattering surface of the $\chi$ particle. The radius $R_{\rm LSS}$ is the value for which a typical thermal $\chi$ particle can escape without undergoing any further scattering.  For the large elastic cross sections of primary interest here, $R_{\rm LSS}$  lies near the surface of the Earth or in the atmosphere,  {\em i.e.}, $R_{\rm LSS} \simeq R_{\oplus}$.

Qualitatively, we find that evaporation is always negligible for DM heavier than 10 GeV, and is always important for $m_\chi \lesssim 1$\,GeV irrespective of the DM-nucleon scattering cross-section~\cite{1990ApJ...356..302G,Garani:2017jcj,Bramante:2022pmn,Garani:2021feo}. Together with the radial distribution $G_\chi(r)$ discussed above, this dictates a mass range over which the direct annihilation of DMCs within the volumes of neutrino telescopes can be observed
\begin{align}
\label{range}
1\,{\rm GeV} \lesssim m_\chi \lesssim 5\,{\rm GeV} \ .
\end{align}
Outside of this mass domain, either $G_{\chi}(R_\oplus)$ or $\tau_{\rm evap}$ is very small and the corresponding annihilation signal is extremely weak. 

Finally, the annihilation rate is given by 
\begin{align}
\tau_{\rm ann}^{-1} &=& \frac{4\pi}{N_\chi^2} \int_0^{R_\oplus}\!dr\,r^2 n^2_\chi(r)\langle \sigma v \rangle_{\rm ann}
\nonumber\\
&\simeq& \frac{4\pi\langle \sigma v \rangle_{\rm ann}}{V_\oplus^2} \int_0^{R_\oplus}\!dr\, r^2 G^2_\chi(r) \ ,
\end{align}
where in the second line we have assumed an approximately constant annihilation cross section $\sigv$, {\em i.e.}, energy-independent $s$-wave annihilation. 

Combining these terms, it is straightforward to integrate Eq.~\eqref{Nchi} and solve for $N_\chi$.  For most of the parameter space relevant for our problem, either the annihilation or evaporation counter-balances the accumulation on timescales $t_{\rm eq}$ shorter than the lifetime of the earth so that $dN_\chi/dt\to 0$. In this case the solution is easily found, $2N_\chi = \left[(\tau_{\rm ann}/\tau_{\rm evap})^2+4\Gamma_{\rm cap}\tau_{\rm ann} \right]^{1/2}-\tau_{\rm ann}/\tau_{\rm evap}  $. 
Depending on the strength of evaporation, two important regimes can be found: 
$N_\chi \simeq \sqrt{\Gamma_{\rm cap}\tau_{\rm ann}}$ when the evaporation is negligible and $N_\chi \simeq \Gamma_{\rm cap} \tau_{\rm evap}$ when it is important. 

\noindent\emph{\textbf{Direct annihilation inside neutrino telescopes:}}
We now compute the annihilation event rate of a DMC within the detector volume of SK
\begin{align}
	\Gamma_{\rm{ann}}^{\rm SK} =\langle \sigma v\rangle_{\rm{ann}} n^2_{\chi} (R_{\oplus})V_{\rm{SK}} = \langle \sigma v\rangle_{\rm{ann}} \frac{N_\chi^2G_{\chi}^2(R_\oplus) V_{\rm{SK}}}{V_\oplus^2}. 
	\label{6}
\end{align}
For this analysis we use the fiducial volume of SK,  $V_{\rm SK} = 2 \times 10^{10}\,\rm{cm}^3$.
If evaporation can be neglected, this reduces to a simple intuitive result,
\begin{align}
	\Gamma_{\rm{ann}}^{\rm{SK}} = \Gamma_{\rm cap} \times \frac{V_{\rm{SK}}G_{\chi}^2(R_\oplus)}{4\pi  \int_0^{R_\oplus} r^2 dr G^2_\chi(r)}\xrightarrow{G_\chi\to 1} \Gamma_{\rm cap} \times \frac{V_{\rm{SK}}}{V_\oplus},
\end{align}
where the second relation applies in the limit of a uniform distribution. For sufficiently large scattering cross sections $\sigma_{\chi n}$ and $m_\chi = 2\,\text{GeV}$, we find annihilation rates in SK of $\Gamma_{\rm ann}^{\rm SK} \simeq 10^6\,\text{yr}^{-1}\,(f_\chi/10^{-5})$ with a DM density of $\simeq  10^5( f_\chi/10^{-5})$ \,GeV\,cm$^{-3}$ at SK's depth [in the limit of zero annihilation, maximal DM density is $\simeq 10^9(f_\chi/10^{-5})$ \,GeV\,cm$^{-3}$]. If the annihilations result in \emph{visible energy}, such rates are very significant, and may even exceed any counting rates in SK by orders of magnitude. We note that this is a drastic departure from the tiny event rate expected for a weakly interacting DM candidate that does not build a large over-concentration near the surface of the Earth~\cite{Undagoitia:2021tza}. 

Given the relevant energy range of annihilations, equal to $m_\chi = 1$--$5$\,GeV, the closest SK experimental analysis for our purposes is the search for di-nucleon decay of Ref.~\cite{Super-Kamiokande:2015jbb,Super-Kamiokande:2018apg}, where the main background is from atmospheric neutrinos. The SK Collaboration has shown that in certain decay channels, such as $nn\to 2\pi^0\to 4\gamma$, cuts on fiducial volume, energy, invariant mass, and multiplicity remove essentially all background, achieving single-event sensitivity~\cite{Super-Kamiokande:2015jbb}. Based on these considerations, we derive an anticipated SK exclusion on annihilating DMCs under the assumptions that the final state allows for a similar background-free identification and can be detected with an efficiency of 10\% as in Ref.~\cite{Super-Kamiokande:2015jbb}. To do so, we compare our predicted detection rates with the limit rate of 3 events for a 282.1 kiloton-yr exposure: $\Gamma_{\rm ann}^{\rm SK} <\Gamma^{\rm SK}_{\rm lim} = 0.24\,\rm yr^{-1}$. While a full experimental analysis is needed, our calculation indicates that new exclusions on the DM-nucleon scattering cross-section could be obtained from existing SK data over the mass range of $m_\chi \simeq 1$--$5$~GeV, even when the annihilating species $\chi$ makes up only a tiny fraction of the DM density. 

We illustrate the anticipated SK sensitivity to DMC annihilation as a function of $\chi$ mass $m_\chi$ and per-nucleon cross section $\sigma_{\chi n}$ in Fig.~\ref{fig1} for $f_{\chi}=10^{-4},\,10^{-6},\,10^{-8}$, and $10^{-10}$. Note that, to make a connection with direct detection constraints, we define an effective per nucleon scattering cross section via  $\sigma_{\chi A} = \sigma_{\chi n}\,A^2 \left(\mu_{\chi\, A}/\mu_{\chi n}\right)^2$ where $A$ is the mass number of the nuclei, and $\mu_{\chi {A}(n)}$ is the reduced mass of the DM-nucleus~(nucleon) system. At the lower end of the DMC mass range, the shapes of the exclusion regions are solely determined  by thermal evaporation, whereas at the upper end they are set by both thermal evaporation and rapid depletion of the surface density of Earth-bound DM due to gravity. Note that the anticipated sensitivity of this method extends down to very tiny DMC fractions. Quantitatively, for  $f_{\chi}=10^{-10}$, $m_\chi = 2.5\,\rm{GeV}$, and $\sigma_{\chi n} = 10^{-28}\,\rm{cm}^2$, the expected event rate at SK can be as high as $15$ events per year, which constitutes a detectable signal. Note as well that the assumption of a background-free search is not entirely crucial for obtaining bounds. Indeed, as Fig.\,\ref{fig1} shows, the change from $f_\chi = 10^{-4}\to 10^{-6}$ leads to a modest reduction of the excluded parameter space at large $m_\chi$. Since the signal is proportional to $f_\chi$, a similar reduction would occur if the experimental limit rate were weakened by a similar factor, $\Gamma^{\rm SK}_{\rm lim} \to 100\times \Gamma^{\rm SK}_{\rm lim}$. We conclude that the  limits from SK are robust, and should be applicable to a wide class of models. 

Also shown in Fig.\,\ref{fig1} for comparison are exclusions from several surface and underground direct detection  searches~\cite{CRESST:2019jnq,CRESST:2017ues,XENON:2018voc,EDELWEISS:2019vjv,DarkSide:2018bpj,Rich:1987st}.  To adjust the experimental bounds given for $f_\chi=1$ to the smaller fractions of interest here, we have applied the simplified method described in Ref.~\cite{McKeen:2022poo}. As shown in Ref.~\cite{Cappiello:2023hza}, this approach gives a reasonable approximation to more computationally intensive calculations such as Refs.~\cite{Emken:2017qmp,Mahdawi:2017cxz,Mahdawi:2017utm,Emken:2018run}. We note, however, that the simplified method we use tends to overestimate slightly the exclusions at small $f_\chi \ll 1$~\cite{Cappiello:2023hza}. Thus, the unexcluded regions where our SK annihilation proposal shows new sensitivity are expected to be robust.

\noindent\emph{\textbf{Secluded Relic Model:}}
To illustrate our results in a concrete model, we consider a dark sector with a Dirac fermion DMC $\chi$ coupled to a dark photon $ A^{\prime}$ with the low-energy effective Lagrangian
\begin{multline}
	\mathcal{L} = -\frac{1}{4} \left(F^{\prime}_{\mu \nu}\right)^2 - \frac{\epsilon}{2} F^{\prime}_{\mu \nu} F^{\mu \nu} + \frac{1}{2} m^2_{A^{\prime}}\left(A^{\prime}_{\mu}\right)^2\\ + \bar{\chi}(i\gamma^{\mu}D_{\mu}-m_{\chi})\chi\,,
\end{multline}
where $\epsilon$ describes kinetic mixing with the photon, $m_{A^{\prime}}$ is the mass of dark photon, $D_{\mu}=\partial_{\mu}-ig_dA^{\prime}_{\mu}$, and $g_d \equiv \sqrt{4\pi\,\alpha_d}$ is the dark coupling constant. 

Annihilation of $\chi$ to dark photons which subsequently decay to SM particles, $\chi \bar{\chi} \to A^{\prime} A^{\prime}$ with $A^{\prime} \to \rm{SM}$, is possible for $m_{A'}<m_\chi$~\cite{Pospelov:2007mp} and efficiently depletes the abundance of $\chi$ to produce $f_\chi \ll 1$ for moderate $\alpha_d$. The annihilation rate during freeze-out can receive a significant non-perturbative enhancement for larger $\alpha_d \gtrsim 0.05$ and $m_\chi \gg m_{A'}$~\cite{Arkani-Hamed:2008hhe,Pospelov:2008jd}. We compute $f_\chi$ in terms of the model parameters assuming thermal freeze-out by approximating the potential between annihilating $\chi$ and $\bar{\chi}$ with a Hulth\`en potential, which has been shown to give a very good estimate of the full result~\cite{Cassel:2009wt,Feng:2010zp}. The perturbative cross section for $\chi$ to scatter on a nucleus $(Z,A)$ is related to the model parameters by~\cite{Pospelov:2007mp}
\begin{align}
\sigma_{\chi A} = \frac{16 \pi Z^2 \alpha \alpha_d \epsilon^2 \mu_{\chi A}^2}{m^4_{A^{\prime}}}\,,
\end{align}
where $Z$ is the atomic number of the nuclei, $m_{A}$ is its mass, and $\alpha$ is the fine-structure constant. 

\begin{figure}[t!]
	\centering
	\includegraphics[width=0.4\textwidth]{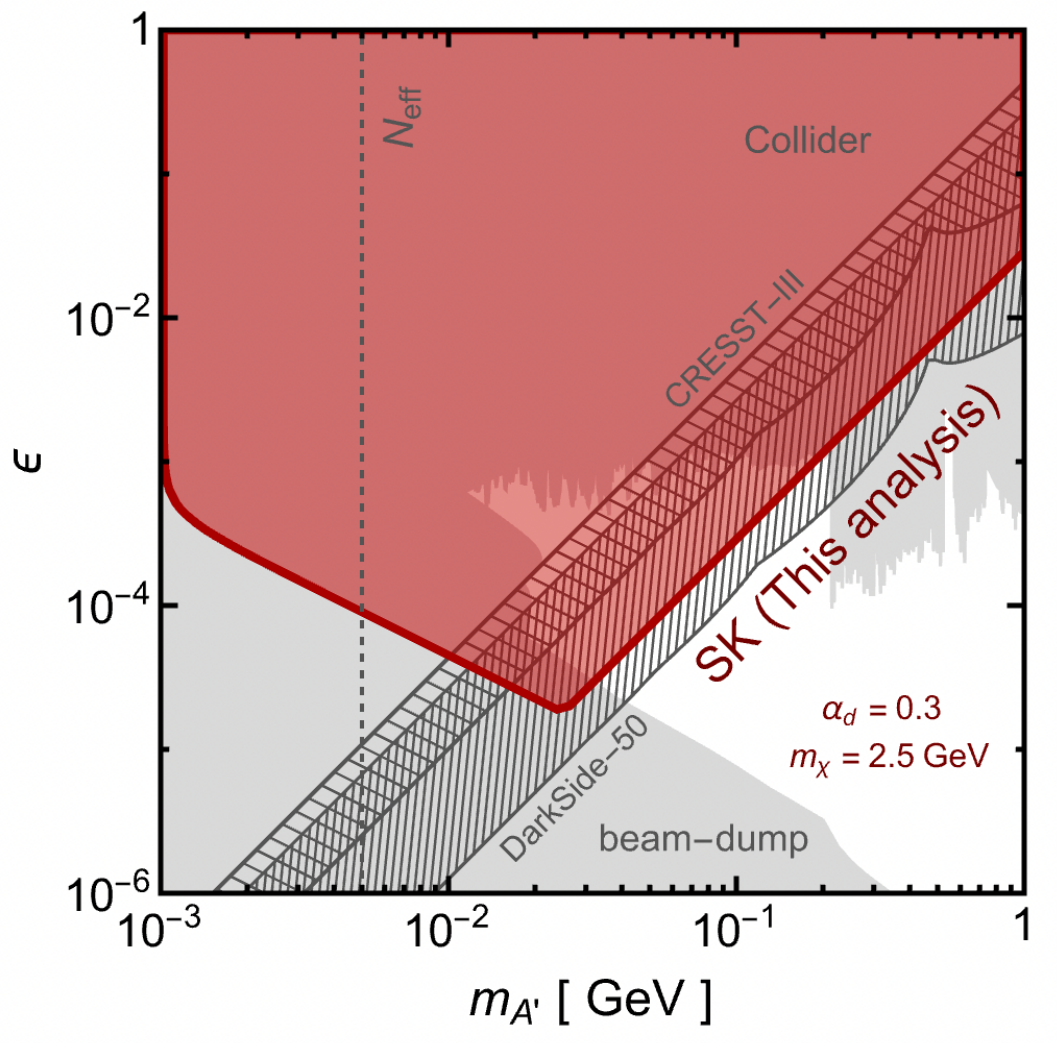}
	\caption{Anticipated sensitivity to a dark photon-mediated DMC $\chi$ for mass $m_\chi = 2.5\,\text{GeV}$, gauge coupling $\alpha_d=0.3$ in terms of the dark photon mass $m_{A^\prime}$ and kinetic mixing $\epsilon$ from annihilation of Earth-bound $\chi$ inside Super Kamiokande~(red shaded). The DM fraction $f_\chi$ of $\chi$ is determined from the model parameters assuming thermal freeze-out in the early Universe. Also shown are bounds from direct DM searches at CRESST~III~\cite{CRESST:2019jnq} and DarkSide-50~\cite{DarkSide:2018bpj}~(gray hatched), as well as searches for a visibly decaying dark photon~\cite{Pospelov:2008zw,Bjorken:2009mm,LHCb:2019vmc,Lanfranchi:2020crw}~(gray shaded).}
\label{fig2}
\end{figure}

In Fig.\,\ref{fig2} we show the sensitivity of our approach to this representative model for $m_\chi = 2.5\,\text{GeV}$ and $\alpha_d = 0.3$ as a function of $m_{A^\prime}$ and $\epsilon$. For these values, the DM fraction of $\chi$ is approximately $f_\chi \simeq 3\times 10^{-9}$, with a mild dependence on $m_{A^\prime}$. The red shaded region in the figure shows the anticipated exclusion from SK, where we apply the same assumptions regarding the experimental sensitivity as before. Note that, for the $A^\prime$ mass range considered the primary dark photon decay modes are to leptons and pions, and are therefore visible and distinctive. In particular, the annihilation process $\chi \bar\chi \to 2 A'\to 2(e^+e^-)$ is very similar in terms of SK signature to $nn\to 2\pi^0\to 4\gamma$ decay~\cite{Super-Kamiokande:2015jbb}. To ensure that the dark photons produced by $\chi\bar\chi$ annihilation decay within the SK fiducial volume, we require further that the SK-frame decay length of the $A^\prime$ is less than 1 m,  {\em i.e.}, $\gamma c\tau_{A^\prime} < 1\,\textrm{m}$; this is important for $m_{A^\prime}\lesssim 20~\rm MeV$. We also show existing bounds on the scenario from direct DM searches~\cite{CRESST:2019jnq,DarkSide:2018bpj},  and from direct searches for a visibly decaying dark photon~\cite{Pospelov:2008zw,Bjorken:2009mm,LHCb:2019vmc,Lanfranchi:2020crw}. The dashed vertical line indicates the lower bound on $m_{A^\prime}$ for a thermalized dark photon from the number of relativistic degrees of freedom during primordial nucleosynthesis in the early Universe~\cite{Krnjaic:2019dzc}.

A final comment is warranted on the possibility of observing the $\chi$ annihilation {\em outside} the Earth's volume using cosmic- and $\gamma$-ray detectors in the GeV range, such as AMS-02~\cite{AMS:2021nhj} and Fermi-LAT~\cite{Fermi-LAT:2009ihh}. By continuity, it is clear that some distribution of $\chi$ (a ``Boltzmanian tail") is present in the atmosphere and above.  Annihilation of $\chi \bar\chi$, with subsequent decay of $A'$ generates electrons, muons, and pions, and therefore contributes to the observed electron and positron flux.
While the counting rates of these experiments are much larger than in SK, there is a gain associated with the fact that the signal is collected from a large volume, for which we take a characteristic orbit height, $h\sim 400\rm\,km$.  The expected additional flux from DM annihilation in the atmosphere, given the SK bound, is 
\begin{align}
\Phi_{\rm ann} \sim \Gamma_{\rm SK}V_{\rm SK}^{-1} \times h < 10^{-10}\,{\rm cm^{-2}s^{-1}}
\end{align}
which is far below the typical electron and positron fluxes measured by the AMS-02 \cite{AMS:2019iwo} that are on the order of $\mathcal{O}(10^{-3}-10^{-2}){\rm\,cm^{-2}\,s^{-1}}$ in this energy range.

\noindent\emph{\textbf{Summary and Conclusion:}}
Earth-bound DM particles can be very abundant near the surface of the Earth if they are sufficiently light and strongly interacting. In this work, we point out that annihilation of an Earth-bound DM component at large underground detectors such as Super-Kamiokande provides a novel technique for their detection. The main strength of this proposal stems from the fact that the energy deposition due to annihilation of Earth-bound DM is not limited by their minuscule amount of kinetic energy, but can instead be as large as their invariant mass, $2m_\chi$. We have demonstrated that this approach can test strongly-interacting DMC over the mass range $m_\chi = 1$--$5$\,GeV down to very small mass fractions, well beyond what is possible with other approaches. The upcoming gigantic underground detectors such as Hyper-Kamiokande~\cite{Hyper-Kamiokande:2018ofw}, JUNO~\cite{JUNO:2021vlw}, DUNE~\cite{DUNE:2020ypp}, and THEIA~\cite{Theia:2019non} will significantly enhance the detection prospects of such Earth-bound DM.

\noindent\emph{\textbf{Acknowledgments:}} 
We thank Christopher Cappiello and Marianne Moore
for helpful discussions. DM and DM are supported by Discovery Grants from the Natural Sciences and Engineering Research Council of Canada (NSERC). TRIUMF receives federal funding via a contribution agreement with the National Research Council (NRC) of Canada.  M.P. is supported in part by U.S. Department of Energy Grant No. DE-SC0011842. MP is grateful to Perimeter Institute for theoretical physics for hospitality. HR is supported in part by NSF Grant PHY-1720397 and the Gordon and Betty Moore Foundation Grant GBMF7946. AR acknowledges support from the National Science Foundation (Grant No. PHY-2020275), and  the Heising-Simons Foundation (Grant 2017-228).  

\bibliographystyle{JHEP}
\bibliography{ref.bib}

\providecommand{\href}[2]{#2}\begingroup\raggedright\begin{thebibliography}{10}

\bibitem{Aghanim:2018eyx}
{\scshape Planck} collaboration, N.~Aghanim et~al., \emph{{Planck 2018 results.
  VI. Cosmological parameters}},
  \href{https://doi.org/10.1051/0004-6361/201833910}{\emph{Astron. Astrophys.}
  {\bfseries 641} (2020) A6}
  [\href{https://arxiv.org/abs/1807.06209}{{\ttfamily 1807.06209}}].

\bibitem{Cooley:2022ufh}
J.~Cooley et~al., \emph{{Report of the Topical Group on Particle Dark Matter
  for Snowmass 2021}},  \href{https://arxiv.org/abs/2209.07426}{{\ttfamily
  2209.07426}}.

\bibitem{Baryakhtar:2022hbu}
M.~Baryakhtar et~al., \emph{{Dark Matter In Extreme Astrophysical
  Environments}},  in \emph{{2022 Snowmass Summer Study}}, 3, 2022,
  \href{https://arxiv.org/abs/2203.07984}{{\ttfamily 2203.07984}}.

\bibitem{Boddy:2022knd}
K.~K. Boddy et~al., \emph{{Snowmass2021 theory frontier white paper:
  Astrophysical and cosmological probes of dark matter}},
  \href{https://doi.org/10.1016/j.jheap.2022.06.005}{\emph{JHEAp} {\bfseries
  35} (2022) 112} [\href{https://arxiv.org/abs/2203.06380}{{\ttfamily
  2203.06380}}].

\bibitem{Farrar:2002ic}
G.~R. Farrar, \emph{{A stable H dibaryon: Dark matter candidate within QCD?}},
  \href{https://doi.org/10.1023/A:1025702431127}{\emph{Int. J. Theor. Phys.}
  {\bfseries 42} (2003) 1211}.

\bibitem{Collar:2018ydf}
J.~I. Collar, \emph{{Search for a nonrelativistic component in the spectrum of
  cosmic rays at Earth}},
  \href{https://doi.org/10.1103/PhysRevD.98.023005}{\emph{Phys. Rev. D}
  {\bfseries 98} (2018) 023005}
  [\href{https://arxiv.org/abs/1805.02646}{{\ttfamily 1805.02646}}].

\bibitem{Zaharijas:2004jv}
G.~Zaharijas and G.~R. Farrar, \emph{{A Window in the dark matter exclusion
  limits}}, \href{https://doi.org/10.1103/PhysRevD.72.083502}{\emph{Phys. Rev.
  D} {\bfseries 72} (2005) 083502}
  [\href{https://arxiv.org/abs/astro-ph/0406531}{{\ttfamily
  astro-ph/0406531}}].

\bibitem{Neufeld:2018slx}
D.~A. Neufeld, G.~R. Farrar and C.~F. McKee, \emph{{Dark Matter that Interacts
  with Baryons: Density Distribution within the Earth and New Constraints on
  the Interaction Cross-section}},
  \href{https://doi.org/10.3847/1538-4357/aad6a4}{\emph{Astrophys. J.}
  {\bfseries 866} (2018) 111}
  [\href{https://arxiv.org/abs/1805.08794}{{\ttfamily 1805.08794}}].

\bibitem{Pospelov:2020ktu}
M.~Pospelov and H.~Ramani, \emph{{Earth-bound millicharge relics}},
  \href{https://doi.org/10.1103/PhysRevD.103.115031}{\emph{Phys. Rev. D}
  {\bfseries 103} (2021) 115031}
  [\href{https://arxiv.org/abs/2012.03957}{{\ttfamily 2012.03957}}].

\bibitem{Leane:2022hkk}
R.~K. Leane and J.~Smirnov, \emph{{Floating Dark Matter in Celestial Bodies}},
  \href{https://arxiv.org/abs/2209.09834}{{\ttfamily 2209.09834}}.

\bibitem{Berlin:2023zpn}
A.~Berlin, H.~Liu, M.~Pospelov and H.~Ramani, \emph{{The Terrestrial Density of
  Strongly-Coupled Relics}},
  \href{https://arxiv.org/abs/2302.06619}{{\ttfamily 2302.06619}}.

\bibitem{Pospelov:2019vuf}
M.~Pospelov, S.~Rajendran and H.~Ramani, \emph{{Metastable Nuclear Isomers as
  Dark Matter Accelerators}},
  \href{https://doi.org/10.1103/PhysRevD.101.055001}{\emph{Phys. Rev. D}
  {\bfseries 101} (2020) 055001}
  [\href{https://arxiv.org/abs/1907.00011}{{\ttfamily 1907.00011}}].

\bibitem{Lehnert:2019tuw}
B.~Lehnert, H.~Ramani, M.~Hult, G.~Lutter, M.~Pospelov, S.~Rajendran et~al.,
  \emph{{Search for Dark Matter Induced Deexcitation of $^{180}$Ta$\rm ^m$}},
  \href{https://doi.org/10.1103/PhysRevLett.124.181802}{\emph{Phys. Rev. Lett.}
  {\bfseries 124} (2020) 181802}
  [\href{https://arxiv.org/abs/1911.07865}{{\ttfamily 1911.07865}}].

\bibitem{McKeen:2022poo}
D.~McKeen, M.~Moore, D.~E. Morrissey, M.~Pospelov and H.~Ramani,
  \emph{{Accelerating Earth-bound dark matter}},
  \href{https://doi.org/10.1103/PhysRevD.106.035011}{\emph{Phys. Rev. D}
  {\bfseries 106} (2022) 035011}
  [\href{https://arxiv.org/abs/2202.08840}{{\ttfamily 2202.08840}}].

\bibitem{Berlin:2021zbv}
A.~Berlin, H.~Liu, M.~Pospelov and H.~Ramani, \emph{{Low-energy signals from
  the formation of dark-matter\textendash{}nucleus bound states}},
  \href{https://doi.org/10.1103/PhysRevD.105.095028}{\emph{Phys. Rev. D}
  {\bfseries 105} (2022) 095028}
  [\href{https://arxiv.org/abs/2110.06217}{{\ttfamily 2110.06217}}].

\bibitem{Budker:2021quh}
D.~Budker, P.~W. Graham, H.~Ramani, F.~Schmidt-Kaler, C.~Smorra and S.~Ulmer,
  \emph{{Millicharged Dark Matter Detection with Ion Traps}},
  \href{https://doi.org/10.1103/PRXQuantum.3.010330}{\emph{PRX Quantum}
  {\bfseries 3} (2022) 010330}
  [\href{https://arxiv.org/abs/2108.05283}{{\ttfamily 2108.05283}}].

\bibitem{Das:2022srn}
A.~Das, N.~Kurinsky and R.~K. Leane, \emph{{Dark Matter Induced Power in
  Quantum Devices}},  \href{https://arxiv.org/abs/2210.09313}{{\ttfamily
  2210.09313}}.

\bibitem{Billard:2022cqd}
J.~Billard, M.~Pyle, S.~Rajendran and H.~Ramani, \emph{{Calorimetric Detection
  of Dark Matter}},  \href{https://arxiv.org/abs/2208.05485}{{\ttfamily
  2208.05485}}.

\bibitem{Pospelov:2007mp}
M.~Pospelov, A.~Ritz and M.~B. Voloshin, \emph{{Secluded WIMP Dark Matter}},
  \href{https://doi.org/10.1016/j.physletb.2008.02.052}{\emph{Phys. Lett. B}
  {\bfseries 662} (2008) 53} [\href{https://arxiv.org/abs/0711.4866}{{\ttfamily
  0711.4866}}].

\bibitem{CRESST:2019jnq}
{\scshape CRESST} collaboration, A.~H. Abdelhameed et~al., \emph{{First results
  from the CRESST-III low-mass dark matter program}},
  \href{https://doi.org/10.1103/PhysRevD.100.102002}{\emph{Phys. Rev. D}
  {\bfseries 100} (2019) 102002}
  [\href{https://arxiv.org/abs/1904.00498}{{\ttfamily 1904.00498}}].

\bibitem{CRESST:2017ues}
{\scshape CRESST} collaboration, G.~Angloher et~al., \emph{{Results on
  MeV-scale dark matter from a gram-scale cryogenic calorimeter operated above
  ground}}, \href{https://doi.org/10.1140/epjc/s10052-017-5223-9}{\emph{Eur.
  Phys. J. C} {\bfseries 77} (2017) 637}
  [\href{https://arxiv.org/abs/1707.06749}{{\ttfamily 1707.06749}}].

\bibitem{XENON:2018voc}
{\scshape XENON} collaboration, E.~Aprile et~al., \emph{{Dark Matter Search
  Results from a One Ton-Year Exposure of XENON1T}},
  \href{https://doi.org/10.1103/PhysRevLett.121.111302}{\emph{Phys. Rev. Lett.}
  {\bfseries 121} (2018) 111302}
  [\href{https://arxiv.org/abs/1805.12562}{{\ttfamily 1805.12562}}].

\bibitem{EDELWEISS:2019vjv}
{\scshape EDELWEISS} collaboration, E.~Armengaud et~al., \emph{{Searching for
  low-mass dark matter particles with a massive Ge bolometer operated
  above-ground}}, \href{https://doi.org/10.1103/PhysRevD.99.082003}{\emph{Phys.
  Rev. D} {\bfseries 99} (2019) 082003}
  [\href{https://arxiv.org/abs/1901.03588}{{\ttfamily 1901.03588}}].

\bibitem{Rich:1987st}
J.~Rich, R.~Rocchia and M.~Spiro, \emph{{A Search for Strongly Interacting Dark
  Matter}}, \href{https://doi.org/10.1016/0370-2693(87)90788-X}{\emph{Phys.
  Lett. B} {\bfseries 194} (1987) 173}.

\bibitem{DarkSide:2018bpj}
{\scshape DarkSide} collaboration, P.~Agnes et~al., \emph{{Low-Mass Dark Matter
  Search with the DarkSide-50 Experiment}},
  \href{https://doi.org/10.1103/PhysRevLett.121.081307}{\emph{Phys. Rev. Lett.}
  {\bfseries 121} (2018) 081307}
  [\href{https://arxiv.org/abs/1802.06994}{{\ttfamily 1802.06994}}].

\bibitem{Bramante:2022pmn}
J.~Bramante, J.~Kumar, G.~Mohlabeng, N.~Raj and N.~Song, \emph{{Light Dark
  Matter Accumulating in Terrestrial Planets: Nuclear Scattering}},
  \href{https://arxiv.org/abs/2210.01812}{{\ttfamily 2210.01812}}.

\bibitem{Gould:1989hm}
A.~Gould and G.~Raffelt, \emph{{Thermal Conduction by Massive Particles}},
  \href{https://doi.org/10.1086/168568}{\emph{Astrophys. J.} {\bfseries 352}
  (1990) 654}.

\bibitem{Dziewonski:1981xy}
A.~M. Dziewonski and D.~L. Anderson, \emph{{Preliminary reference earth
  model}}, \href{https://doi.org/10.1016/0031-9201(81)90046-7}{\emph{Phys.
  Earth Planet. Interiors} {\bfseries 25} (1981) 297}.

\bibitem{https://doi.org/10.1002/2017JB014723}
Y.~Zhang, T.~Sekine, J.-F. Lin, H.~He, F.~Liu, M.~Zhang et~al., \emph{Shock
  compression and melting of an fe-ni-si alloy: Implications for the
  temperature profile of the earth's core and the heat flux across the
  core-mantle boundary},
  \href{https://doi.org/https://doi.org/10.1002/2017JB014723}{\emph{Journal of
  Geophysical Research: Solid Earth} {\bfseries 123} (2018) 1314}.

\bibitem{1990ApJ...356..302G}
A.~{Gould}, \emph{{Evaporation of WIMPs with Arbitrary Cross Sections}},
  \href{https://doi.org/10.1086/168840}{\emph{Astrophysical Journal} {\bfseries
  356} (1990) 302}.

\bibitem{Garani:2017jcj}
R.~Garani and S.~Palomares-Ruiz, \emph{{Dark matter in the Sun: scattering off
  electrons vs nucleons}},
  \href{https://doi.org/10.1088/1475-7516/2017/05/007}{\emph{JCAP} {\bfseries
  05} (2017) 007} [\href{https://arxiv.org/abs/1702.02768}{{\ttfamily
  1702.02768}}].

\bibitem{Garani:2021feo}
R.~Garani and S.~Palomares-Ruiz, \emph{{Evaporation of dark matter from
  celestial bodies}},  \href{https://arxiv.org/abs/2104.12757}{{\ttfamily
  2104.12757}}.

\bibitem{Undagoitia:2021tza}
T.~M. Undagoitia, W.~Rodejohann, T.~Wolf and C.~E. Yaguna, \emph{{Laboratory
  limits on the annihilation or decay of dark matter particles}},
  \href{https://doi.org/10.1093/ptep/ptab139}{\emph{PTEP} {\bfseries 2022}
  (2022) 013F01} [\href{https://arxiv.org/abs/2107.05685}{{\ttfamily
  2107.05685}}].

\bibitem{Super-Kamiokande:2015jbb}
{\scshape Super-Kamiokande} collaboration, J.~Gustafson et~al., \emph{{Search
  for dinucleon decay into pions at Super-Kamiokande}},
  \href{https://doi.org/10.1103/PhysRevD.91.072009}{\emph{Phys. Rev. D}
  {\bfseries 91} (2015) 072009}
  [\href{https://arxiv.org/abs/1504.01041}{{\ttfamily 1504.01041}}].

\bibitem{Super-Kamiokande:2018apg}
{\scshape Super-Kamiokande} collaboration, S.~Sussman et~al., \emph{{Dinucleon
  and Nucleon Decay to Two-Body Final States with no Hadrons in
  Super-Kamiokande}},  \href{https://arxiv.org/abs/1811.12430}{{\ttfamily
  1811.12430}}.

\bibitem{Cappiello:2023hza}
C.~Cappiello, \emph{{An Analytic Approach to Light Dark Matter Propagation}},
  \href{https://arxiv.org/abs/2301.07728}{{\ttfamily 2301.07728}}.

\bibitem{Emken:2017qmp}
T.~Emken and C.~Kouvaris, \emph{{DaMaSCUS: The Impact of Underground
  Scatterings on Direct Detection of Light Dark Matter}},
  \href{https://doi.org/10.1088/1475-7516/2017/10/031}{\emph{JCAP} {\bfseries
  10} (2017) 031} [\href{https://arxiv.org/abs/1706.02249}{{\ttfamily
  1706.02249}}].

\bibitem{Mahdawi:2017cxz}
M.~S. Mahdawi and G.~R. Farrar, \emph{{Closing the window on $\sim$GeV Dark
  Matter with moderate ($\sim$ $\mu$b) interaction with nucleons}},
  \href{https://doi.org/10.1088/1475-7516/2017/12/004}{\emph{JCAP} {\bfseries
  12} (2017) 004} [\href{https://arxiv.org/abs/1709.00430}{{\ttfamily
  1709.00430}}].

\bibitem{Mahdawi:2017utm}
M.~S. Mahdawi and G.~R. Farrar, \emph{{Energy loss during Dark Matter
  propagation in an overburden}},
  \href{https://arxiv.org/abs/1712.01170}{{\ttfamily 1712.01170}}.

\bibitem{Emken:2018run}
T.~Emken and C.~Kouvaris, \emph{{How blind are underground and surface
  detectors to strongly interacting Dark Matter?}},
  \href{https://doi.org/10.1103/PhysRevD.97.115047}{\emph{Phys. Rev. D}
  {\bfseries 97} (2018) 115047}
  [\href{https://arxiv.org/abs/1802.04764}{{\ttfamily 1802.04764}}].

\bibitem{Arkani-Hamed:2008hhe}
N.~Arkani-Hamed, D.~P. Finkbeiner, T.~R. Slatyer and N.~Weiner, \emph{{A Theory
  of Dark Matter}},
  \href{https://doi.org/10.1103/PhysRevD.79.015014}{\emph{Phys. Rev. D}
  {\bfseries 79} (2009) 015014}
  [\href{https://arxiv.org/abs/0810.0713}{{\ttfamily 0810.0713}}].

\bibitem{Pospelov:2008jd}
M.~Pospelov and A.~Ritz, \emph{{Astrophysical Signatures of Secluded Dark
  Matter}}, \href{https://doi.org/10.1016/j.physletb.2008.12.012}{\emph{Phys.
  Lett. B} {\bfseries 671} (2009) 391}
  [\href{https://arxiv.org/abs/0810.1502}{{\ttfamily 0810.1502}}].

\bibitem{Cassel:2009wt}
S.~Cassel, \emph{{Sommerfeld factor for arbitrary partial wave processes}},
  \href{https://doi.org/10.1088/0954-3899/37/10/105009}{\emph{J. Phys. G}
  {\bfseries 37} (2010) 105009}
  [\href{https://arxiv.org/abs/0903.5307}{{\ttfamily 0903.5307}}].

\bibitem{Feng:2010zp}
J.~L. Feng, M.~Kaplinghat and H.-B. Yu, \emph{{Sommerfeld Enhancements for
  Thermal Relic Dark Matter}},
  \href{https://doi.org/10.1103/PhysRevD.82.083525}{\emph{Phys. Rev. D}
  {\bfseries 82} (2010) 083525}
  [\href{https://arxiv.org/abs/1005.4678}{{\ttfamily 1005.4678}}].

\bibitem{Pospelov:2008zw}
M.~Pospelov, \emph{{Secluded U(1) below the weak scale}},
  \href{https://doi.org/10.1103/PhysRevD.80.095002}{\emph{Phys. Rev. D}
  {\bfseries 80} (2009) 095002}
  [\href{https://arxiv.org/abs/0811.1030}{{\ttfamily 0811.1030}}].

\bibitem{Bjorken:2009mm}
J.~D. Bjorken, R.~Essig, P.~Schuster and N.~Toro, \emph{{New Fixed-Target
  Experiments to Search for Dark Gauge Forces}},
  \href{https://doi.org/10.1103/PhysRevD.80.075018}{\emph{Phys. Rev. D}
  {\bfseries 80} (2009) 075018}
  [\href{https://arxiv.org/abs/0906.0580}{{\ttfamily 0906.0580}}].

\bibitem{LHCb:2019vmc}
{\scshape LHCb} collaboration, R.~Aaij et~al., \emph{{Search for
  $A'\to\mu^+\mu^-$ Decays}},
  \href{https://doi.org/10.1103/PhysRevLett.124.041801}{\emph{Phys. Rev. Lett.}
  {\bfseries 124} (2020) 041801}
  [\href{https://arxiv.org/abs/1910.06926}{{\ttfamily 1910.06926}}].

\bibitem{Lanfranchi:2020crw}
G.~Lanfranchi, M.~Pospelov and P.~Schuster, \emph{{The Search for Feebly
  Interacting Particles}},
  \href{https://doi.org/10.1146/annurev-nucl-102419-055056}{\emph{Ann. Rev.
  Nucl. Part. Sci.} {\bfseries 71} (2021) 279}
  [\href{https://arxiv.org/abs/2011.02157}{{\ttfamily 2011.02157}}].

\bibitem{Krnjaic:2019dzc}
G.~Krnjaic and S.~D. McDermott, \emph{{Implications of BBN Bounds for Cosmic
  Ray Upscattered Dark Matter}},
  \href{https://doi.org/10.1103/PhysRevD.101.123022}{\emph{Phys. Rev. D}
  {\bfseries 101} (2020) 123022}
  [\href{https://arxiv.org/abs/1908.00007}{{\ttfamily 1908.00007}}].

\bibitem{AMS:2021nhj}
{\scshape AMS} collaboration, M.~Aguilar et~al., \emph{{The Alpha Magnetic
  Spectrometer (AMS) on the international space station: Part II \textemdash{}
  Results from the first seven years}},
  \href{https://doi.org/10.1016/j.physrep.2020.09.003}{\emph{Phys. Rept.}
  {\bfseries 894} (2021) 1}.

\bibitem{Fermi-LAT:2009ihh}
{\scshape Fermi-LAT} collaboration, W.~B. Atwood et~al., \emph{{The Large Area
  Telescope on the Fermi Gamma-ray Space Telescope Mission}},
  \href{https://doi.org/10.1088/0004-637X/697/2/1071}{\emph{Astrophys. J.}
  {\bfseries 697} (2009) 1071}
  [\href{https://arxiv.org/abs/0902.1089}{{\ttfamily 0902.1089}}].

\bibitem{AMS:2019iwo}
{\scshape AMS} collaboration, M.~Aguilar et~al., \emph{{Towards Understanding
  the Origin of Cosmic-Ray Electrons}},
  \href{https://doi.org/10.1103/PhysRevLett.122.101101}{\emph{Phys. Rev. Lett.}
  {\bfseries 122} (2019) 101101}.

\bibitem{Hyper-Kamiokande:2018ofw}
{\scshape Hyper-Kamiokande} collaboration, K.~Abe et~al.,
  \emph{{Hyper-Kamiokande Design Report}},
  \href{https://arxiv.org/abs/1805.04163}{{\ttfamily 1805.04163}}.

\bibitem{JUNO:2021vlw}
{\scshape JUNO} collaboration, A.~Abusleme et~al., \emph{{JUNO physics and
  detector}}, \href{https://doi.org/10.1016/j.ppnp.2021.103927}{\emph{Prog.
  Part. Nucl. Phys.} {\bfseries 123} (2022) 103927}
  [\href{https://arxiv.org/abs/2104.02565}{{\ttfamily 2104.02565}}].

\bibitem{DUNE:2020ypp}
{\scshape DUNE} collaboration, B.~Abi et~al., \emph{{Deep Underground Neutrino
  Experiment (DUNE), Far Detector Technical Design Report, Volume II: DUNE
  Physics}},  \href{https://arxiv.org/abs/2002.03005}{{\ttfamily 2002.03005}}.

\bibitem{Theia:2019non}
{\scshape Theia} collaboration, M.~Askins et~al., \emph{{THEIA: an advanced
  optical neutrino detector}},
  \href{https://doi.org/10.1140/epjc/s10052-020-7977-8}{\emph{Eur. Phys. J. C}
  {\bfseries 80} (2020) 416}
  [\href{https://arxiv.org/abs/1911.03501}{{\ttfamily 1911.03501}}].

\end{thebibliography}\endgroup
\end{document}